\documentclass[12pt,a4paper]{article}
\usepackage{epsfig}
\title{Production of the neutral toppion at the $e\gamma$ colliders }

\author{Xuelei Wang $^{(a,b)}$, Yueling Yang$^{b}$,Bingzhong
Li$^{b}$, Lingde Wan$^{b}$\\ {\small a: CCAST(World Laboratory)
P.O. BOX 8730. B.J. 100080P.R.China}\\ {\small b: College of
Physics and Information Engineering,}\\ \small{Henan Normal
University, Xinxiang  453002. P.R.China}
\thanks{This work is supported by the National Natural Science
Foundation of China, the Excellent Youth Foundation of Henan
Scientific Committee, the Henan Innovation Project for University
Prominent Research Talents.}
\thanks{E-mail:wangxuelei@263.net}
\thanks{Mailing address} }
\begin{document}
\maketitle

\begin{abstract}
\hspace{5mm} In the framework of topcolor-assisted
technicolor(TC2) model, we study a neutral toppion production
process $e^{-}\gamma\rightarrow e^{-}\Pi^{0}_{t}$ in this paper.
Our results show that the production cross section of
$e^{-}\gamma\rightarrow e^{-}\Pi^{0}_{t}$ can reach the level of
several tens fb, and
 over $10^{3}$ neutral toppion events can be produced in the
planned $e^+e^-$ linear colliders each year. Therefore, such a
toppion production process provides us a unique chance to detect
toppion events and test the TC2 model. On the other hand, the
cross section of $e^{-}\gamma\rightarrow e^{-}\Pi^{0}_{t}$ is
about one order of magnitude larger than those of some similar
processes in SM and MSSM(i.e., $e^{-}\gamma\rightarrow e^{-}H$ in
SM and $e^{-}\gamma\rightarrow e^{-}H^{0}(A^0,h^0)$ in MSSM). So,
we can easily distinguish the neutral toppion from other neutral
Higgs bosons in SM and MSSM.
\end {abstract}

\vspace{1.0cm} \noindent {\bf PACS number(s)}: 12.60Nz, 14.80.Mz,
12.15.LK, 14.65.Ha

\newpage
\noindent{\bf I. Introduction}~~\\
Although the
Glashow-Weinberg-Salam(GWS)theory which bases on the gauge group
$SU_{L}(2)\otimes U_{Y}(1)$ have made a great success to describe
the weak and electromagnetic interactions, the  mechanism of the
electroweak symmetry breaking(EWSB) is still unknown. So probing
the mechanism of the EWSB will not only be one of the main
subjects of theoretical research but also be the most important
task at future high energy colliders.

Dynamical EWSB, such as technicolor(TC) theory \cite{TC}, is
   an attractive idea that avoids the shortcoming of triviality
   and unnaturalness arising from the elementary Higgs field in
   the standard model(SM). The simplest QCD-like TC
   models\cite{QCD-like} leads to a large oblique correction to
   the electroweak parameter S\cite{s-parameter} and is already
   ruled out by the CERN $e^{+}e^{-}$ collider LEP precision
   electroweak measurement data\cite{LEP1,LEP2}. Various improvements
   have been made to make the predictions consistent
with the LEP precision measurement data. Among all these improved
TC models, topcolor-assisted technicolor(TC2) model \cite{TC2}
   is a more realistic one, which provides an additional source of
   EWSB and also solves heavy top quark problem. In TC2 theory, the new
    strong dynamics topcolor is assumed to be chiral critically strong
    at the scale 1 TeV, and it is coupled preferentially to the third generation.
   In this model, the EWSB is driven mainly by TC interactions and extended technicolor
   gives the contributions to
   all ordinary quark and lepton masses including a very small
   portion of the top quark masses :$m_{t}^{'}=\varepsilon m_{t}
   (0.03\leq\varepsilon\leq0.1)$ \cite{Buchalla}. The topcolor
   interactions also make small contributions to the EWSB and give
   rise to the main part of the top mass $(1-\varepsilon)m_{t}$.
   Three Pseudo-Goldstone bosons (PGB's) called  toppions
   $\Pi^{0}_{t},\Pi^{\pm}_{t}$ are predicted by TC2 model in the
   few hundred GeV region. The physical particle toppions can be
   regarded as the typical feature of TC2 model. Thus, the studies
   of some toppion production processes at present and future high
   energy colliders can help the experiment to search for toppion and test
   TC2 theory, furthermore, to probe EWSB mechanism. A
   comprehensive review on the phenomenological studies in TC2
   model has been given in Ref.\cite{review}.

 Over the last decade, several laboratories in the world have been working on linear $e^+e^-$
 collider projects with an energy from several hundreds GeV up to several TeV and the
luminosity over 100 $fb^{-1}$/year, these are NLC(USA)
 \cite{NLC},JLC(Japan)\cite{JLC}, TESLA(Europe)\cite{TESLA}.
 The search for Higgs particle in SM or some new particles predicted in the
 models beyond the SM(such as Higgs bosons $A^{0},H^{0},h^{0},H^{\pm}$ in MSSM and PGB's
  in TC model) is
  one of the most important goals of future high energy $e^{+}e^{-}$
  colliders. Some Higgs bosons production processes in SM and MSSM in
  $e^{+}e^{-}$ collision have been studied in many literatures\cite{eeH}.
 To search for the toppions in TC2 model,
   the literatures have studied the neutral toppion production processes
 in high energy $e^{+}e^{-}$ collision\cite{yue,wang}. Ref.\cite{yue} has calculated the
   production cross sections of the processes
   $e^{+}e^{-}\to\Pi^{0}_{t}\gamma,\Pi^{0}_{t}Z$ and the results show
   that the cross sections are about several fb. Recently, we have
   studied a flavor-changing neutral toppion production process $e^{+}e^{-}\to
   t\bar{c}\Pi^{0}_{t}$\cite{wang}. We find that the resonance effect
   can enhance the cross section significantly when toppion mass is
   small. The above studies provide the feasible ways to detect
   toppion events and test TC2 model.
The future $e^+e^-$ colliders can also operate in the $e\gamma$ or
 $\gamma\gamma$ modes. High energy photons for $\gamma \gamma, e\gamma$ collisions
can be obtained using compton backscattering of laser light off
the high energy electrons. In this case, the energy and luminosity
of the photon beam would be the same order of
 magnitude of the parent electron beam and the set of final states
 at a photon collider is much richer than that at in an $e^{+}e^{-}$
 mode. At the same time, the high energy photons polarizations
 can relatively easily vary, which is  advantageous for
 experiments. All the virtues of the photon colliders will provide
 us a good  chance to pursuit new physics particles. The production of Higgs bosons
in SM and MSSM at $e\gamma$ colliders have been studied in Ref.\cite{gamma-e}.

   In this paper, in the framework of TC2, we will study a neutral toppion production
   process $e^{-}\gamma\rightarrow e^{-}\Pi^{0}_{t}$. The results show that
 the cross section can be up to the level of several tens fb due to strong coupling of
$\Pi^0_t$ to $t\bar{t}$ and t-channel effect. The signals of
toppion can be easily detected at $e\gamma$ colliders. On the
other hand, we find that we can distinguish toppion from other
toppion-like particles(such as Higgs bosons in SM and MSSM).

\noindent{\bf II. The cross section of the process}~~\\
  As it is known, the couplings of toppions to the three family
  fermions are non-universal and the toppions have large Yukawa couplings
  to the third generation. The coupling of the neutral toppion $\Pi_{t}^{0}$
   to a pair of top quarks is proportion to the mass of top quark
   and the explicit form can be written as\cite{He}:
  \begin{eqnarray}
  i\frac{m_{t}}{\upsilon _{w}}\tan\beta K_{UR}^{tt}K_{UL}^{tt*}
  \overline{t}\gamma_5 t\Pi_{t}^{0}
\end{eqnarray}
 where
 $\tan\beta=\sqrt{(\frac{\upsilon_{\omega}}{\upsilon_{t}})^{2}-1}$
 \hspace{0.5cm} $\upsilon_{\omega}=246$ GeV is electroweak
 symmetry-breaking scale, and $\upsilon_{t}\approx 60-100$ GeV is the
 toppion decay constant. $K^{tt}_{UL}$ is the matrix
  element of the unitary matrix $K_{UL}$ which the CKM matrix
  can be derived as $V=K^{-1}_{UL}K_{DL}$ and $K^{ij}_{UR}$ are
  the matrix
  elements of right-handed rotation matrix  $K_{UR}$. Their values
  can be taken as:
  \begin{eqnarray*}
  \hspace*{1cm}K^{tt*}_{UL}\approx 1 \hspace{1.5cm}K^{tt}_{UR}=1-\varepsilon
\end{eqnarray*}
  Here we take the parameter $\varepsilon$ as a free parameter
  changing from 0.03 to 0.1.

   With $\Pi^0_tt\bar{t}$ coupling, the neutral toppion $\Pi^{0}_{t}$,
   as an isospin-triplet, can
   couple to a pair of gauge bosons through the top quark triangle
   loop in an isospin violating way. Calculating the top quark triangle loop, we can
   explicitly  obtain the couplings of $\Pi^{0}_{t}-\gamma-\gamma$ and
   $\Pi^{0}_{t}-\gamma-Z$
   \begin{eqnarray}
   \Pi^{0}_{t}-\gamma-\gamma \hspace{1.5cm} iN_{c}\frac{8}{9\pi}
   \frac{tan\beta}{\upsilon_{w}}m_{t}^{2}(1-\varepsilon)\alpha_{e}
   \varepsilon_{\mu\nu\rho\delta}p_{2}^{\rho}p_{4}^{\delta}C_{0}
   \end{eqnarray}
\begin{eqnarray}
  \Pi^{0}_{t}-\gamma-Z \hspace{1.5cm} iN_{c}
  \frac{\alpha_e}{3\pi c_ws_w}
   \frac{tan\beta}{\upsilon_{w}}m_{t}^{2}(1-\varepsilon)
   \varepsilon_{\mu\nu\rho\delta}(1-\frac{8}{3}s_{w}^{2})
   p_{2}^{\rho}p_{4}^{\delta}C_{0}
\end{eqnarray}
   where $N_{c}$ is the color index with $N_{c}=3$, $s_{w}=\sin
   \theta_{w}$, $c_{w}=\cos\theta_{w}$($\theta_{w}$
   is the Weinberg angle),
   $C_{0}=C_{0}(-p_{2},p_{4},m_{t},m_{t},m_{t})$ is
    standard three-point scalar integral with
   $p_{2}$ and $p_{4}$ denoting the momenta of the incoming photon
   and the outcoming toppion, respectively.

   With the couplings of $\Pi^0_t\gamma\gamma$ and $\Pi^0_tZ\gamma$, the neutral
   toppion can be produced via the process $e^{-}\gamma\rightarrow e^{-}
   \Pi^{0}_{t}$, the Feynman diagram of the process is shown in Fig.1.
The amplitude of the process can be written directly
\begin{eqnarray}
M=M^{\gamma}+M^{Z}
\end{eqnarray}

\begin{eqnarray}
 M^{\gamma} &=& -iN_{c}\frac{16\sqrt{\pi}}{9\pi}\frac{\tan\beta}{\upsilon_{w}}
 m^{2}_{t}(1-\varepsilon)\alpha_{e}^{3/2}C_{0}
\varepsilon^{\mu\nu\rho\delta}p_{2\rho}p_{4\delta}\\ \nonumber
&&\epsilon_{\mu}(p_{2})
\overline{u}_{e}(p_{3})\gamma_{\nu}u_{e}(p_{1})G(p_{2}-p_{4},0)
\end{eqnarray}

\begin{eqnarray}
 M^{Z} &=& iN_{c}\frac{2\alpha^{3/2}_e}{3\sqrt{\pi}c^2_ws^2_w}
 \frac{\tan\beta}{\upsilon_{w}}(1-\varepsilon)m^{2}_{t}
 (1-\frac{8}{3}s_{w}^{2})C_{0}\\ \nonumber& &
 \varepsilon^{\mu\nu\rho\delta}p_{2\rho}p_{4\delta}\epsilon_{\mu}(p_{2})
\overline{u}_{e}(p_{3})[-\frac{1}{2}L+s^{2}_{w}]\gamma_{\nu}u_{e}(p_{1})
\\ \nonumber &&
 G(p_{2}-p_{4},M_{Z})
\end{eqnarray}
Where, $L=\frac{1}{2}(1-\gamma_{5})$.
$G(p,m)=\frac{1}{p^{2}-m^{2}}$ denotes the propagator of the
particle. We can see that there exists a t-channel resonance
effect for photon, this t-channel resonance effect will enhance
the cross section significantly.

The hard photon beam of the $e\gamma$ collider can be obtained
from laser backscattering at the $e^+e^-$ linear collider. Let
$\hat{s}$ and $s$ be the center-of-mass energies of the $e\gamma$
and $e^+e^-$ systems, respectively. After calculating the cross
section $\sigma(\hat{s})$ for the subprocess $e^-\gamma\to
e^-\Pi^0_t$, the total cross section at the $e^+e^-$ linear
collider can be obtained by folding $\sigma(\hat{s})$ with the
photon distribution function which is given in Ref\cite{Gjikia}
\begin{eqnarray*}                                                   
\sigma_{tot}=\int\limits_{M_{\Pi}^2/s}^{x_{max}}dx\hat{\sigma}
(\hat{s})f_{\gamma}(x)\,,
\end{eqnarray*}
where
\begin{eqnarray*}                                                    
\displaystyle f_\gamma(x)=\frac{1}{D(\xi)}\left[1-x+\frac{1}{1-x}
-\frac{4x}{\xi(1-x)}+\frac{4x^2}{\xi^2(1-x)^2}\right],
\end{eqnarray*}
with
\begin{eqnarray*}                                                    
\displaystyle D(\xi)=\left(1-\frac{4}{\xi}-\frac{8}{\xi^2}\right)
\ln(1+\xi)+\frac{1}{2}+\frac{8}{\xi}-\frac{1}{2(1+\xi)^2}\,.
\end{eqnarray*}
In above equation, $\xi=4E_e\omega_0/m_e^2$ in which $m_e$ and
$E_e$ stand, respectively, for the incident electron mass and
energy, $\omega_0$ stands for the laser photon energy, and
$x=\omega/E_e$ stands for the fraction of energy of the incident
electron carried by the back-scattered photon. $f_\gamma$ vanishes
for $x>x_{max} =\omega_{max}/E_e=\xi/(1+\xi)$. In order to avoid
the creation of $e^+e^-$ pairs by the interaction of the incident
and back-scattered photons, we require $\omega_0x_{max}\leq
m_e^2/E_e$ which implies that $\xi\leq 2+ 2\sqrt{2}\approx 4.8$.
For the choice of $\xi=4.8$, we obtain
\begin{eqnarray*}                                                     
x_{max}\approx 0.83,\hspace{1cm} D(\xi)\approx 1.8 \,.
\end{eqnarray*}
For simplicity, we have ignored the possible polarization for the
electron and photon beams.
 \noindent{\bf III. The numberal results
and conclusions }~~ \\ To obtain numerical results , we take
$m_{t}=174$ GeV, $M_Z=91.187$ GeV, $\upsilon_{t}=60$ GeV,
 $s^2_w=0.23$. The
electromagnetic fine structure constant $\alpha_e$ at certain
energy scale is calculated from the simple QED one-loop evolution
formula with the boundary value $\alpha_e=1/137.04$
\cite{Donoghue}. There are three free parameters in the cross
section, i.e., $\varepsilon,M_{\Pi},s$. To see the influence of
these parameters on the cross section, we take the mass of toppion
$M_{\Pi}$ to vary in certain range 150 GeV$\leq M_{\Pi}\leq 450$
GeV, $\varepsilon=0.03,0.06,0.1$, respectively. Considering the
 center-of-mass energies $\sqrt{s}$ in planned $e^+e^-$ linear colliders(for example:
 TESLA), we take $\sqrt{s}$=500 GeV, 800 GeV, 1600
 GeV, respectively. The final
numerical results of the cross section are summarized in Fig.2-4.
The Fig.2-4 are the plots of the cross section as the function of
$M_{\Pi}$ for $\sqrt{s}$=500 GeV, 800 GeV, 1600 GeV, respectively.
We can see that there is a peak in the plot when $M_{\Pi}$ is
about 350 GeV which arises from top quark triangle loop. We can
see that the cross section is in the range of a few tens fb. With
the luminosity of 100 $fb^{-1}$/year, there are over $10^3$ events
of neutral toppion to be produced via the process $e^-\gamma
\rightarrow e^-\Pi^0_t$ per year. Such sufficient events can be
easily detected experimentally. As have been studied in
Ref\cite{yue,wang}, the cross sections of neutral toppion in
$e^+e^-$ collision are only at the level of a few fb. The
t-channel resonance effect can enhance the cross section of the
process $e^-\gamma\rightarrow e^-\Pi^0_t$ significantly, this
makes process $e^-\gamma\rightarrow e^-\Pi^0_t$ potentially
important for the detecting of toppion. Some Higgs bosons
production processes in $e\gamma$ collision have been studied in
SM and MSSM($e^-\gamma \rightarrow e^-H^0$ in SM and $e^-\gamma
\rightarrow e^-H^0(h^0,A^0)$ in MSSM)\cite{gamma-e}, the results
show that the cross sections are at the level of a few fb, i.e.,
the cross section of $e^{-}\gamma\rightarrow e^{-}\Pi^{0}_{t}$ is
about one order of magnitude larger than those of some similar
processes in SM and MSSM. The reason is that there is a large
extra coefficient $tan\beta$ in the coupling $\Pi^0_tt\bar{t}$
compared with the coupling $Ht\bar{t}$ in SM(MSSM) and $tan\beta$
can enhance the cross section about one order of magnitude. With
such a large cross section of $e^-\gamma \rightarrow e^-\Pi^0_t$,
we can easily distinguish neutral toppion in TC2 from Higgs bosons
in SM and MSSM. This is another important feature of the process
$e^-\gamma\rightarrow e^-\Pi^0_t$.

To determine which channel is the best one to search for neutral
toppion, we need to know its decay branching ratio of each decay
modes. The possible decay modes are: $t\bar{t}$(if
$\Pi^0_t>2m_t$),$t\bar{c}, b\bar{b},gg,\gamma\gamma,Z\gamma$. For
$\Pi^0_t>2m_t$, the main decay mode is $\Pi^0_t \rightarrow
t\bar{t}$. The decay branching ratio $Br(\Pi^0_t\rightarrow
t\bar{c})$ is the largest one when $t\bar{t}$ channel is
forbidden. Using the FormCalc\cite{FormCalc}, we can directly
obtain the cross section of the processes: $e^-\gamma \rightarrow
e^-t\bar{t},e^-\gamma \rightarrow e^-t\bar{c},e^-\gamma
\rightarrow e^-b\bar{b}$ in SM, the results are shown in Table 1.

\null\noindent ~~{\bf Tbale 1:} The cross section of $e^-\gamma
\rightarrow e^-t\bar{t},e^-\gamma \rightarrow
e^-t\bar{c},e^-\gamma \rightarrow e^-b\bar{b}$ in SM.
\vspace{0.1in}
\begin{center}
\doublerulesep 0.8pt \tabcolsep 0.1in
\begin{tabular}{||c|c|c|c||}\hline\hline
$\sqrt{s}(GeV)$ &$\sigma(e^-\gamma \rightarrow e^-t\bar{t})$(pb)&
 $\sigma(e^-\gamma\rightarrow e^-t\bar{c})(pb)$&
$\sigma(e^-\gamma \rightarrow e^-b\bar{b})$(pb)\\ \hline
500&$1.0\times10^{-2}$&$5.5\times 10^{-12}$&10.5\\ \hline
800&$2.7\times10^{-2}$&$7.2\times 10^{-12}$&10.9\\ \hline
1600&$4.2\times 10^{-2}$&$8.6\times 10^{-12}$&11.5
\\ \hline\hline
\end{tabular}
\end {center}

\vspace{0.5cm}

We can see that, in SM,  the cross section of
$e^-\gamma\rightarrow e^-t\bar{c}$ is very small because there is
no tree level flavor-changing neutral current(FCNC) in SM.
Therefore, $e^-\gamma\rightarrow e^-\Pi^0_t \rightarrow
e^-t\bar{c}$ is the most ideal channel to detect neutral toppion.
The decay branching ratio of $\Pi^0_t \rightarrow t\bar{c}$ and
the signal per year in the $t\bar{c}$ channel are shown in table 2

\null\noindent ~~{\bf Table 2:}The decay branching ratio of
$\Pi^0_t \rightarrow t\bar{c}$ and the signal per year in the
$t\bar{c}$ channel. We take $\varepsilon=0.06$ and the luminosity
$L=100 fb^{-1}/$year.  \vspace{0.1in}
\begin{center}
\doublerulesep 0.8pt \tabcolsep 0.1in
\begin{tabular}{||c|c|c|c|c|c|c||}\hline\hline
$M_{\Pi}(GeV)$&\multicolumn{3}{|c|}{160}&\multicolumn{3}{|c|}{400}\\
\hline $Br(\Pi^0_t \rightarrow
t\bar{c})$&\multicolumn{3}{|c|}{0.66}&\multicolumn{3}{|c|}{0.08}\\
\hline $\sqrt{s}(GeV)$&500&800&1600&500&800&1600\\ \hline
Signal/Year in tc channel &600&704&741&73&184&269
\\ \hline\hline
\end{tabular}
\end {center}

\vspace{0.5cm}

We can conclude that there is about a few hundred signals of
neutral toppion  produced in $t\bar{c}$ channel. With such large
numbers of signals and very clean background in SM for this
$t\bar{c}$ channel(As it is shown in table 2 that the cross
section of $e^-\gamma \rightarrow e^-t\bar{c}$ in SM is only about
$10^{-12}$ pb), the neutral toppion can be easily detected via
$t\bar{c}$ channel at $e\gamma$ collision.

 In conclusion, we have studied a neutral toppion production
 process $e^{-}\gamma\rightarrow e^{-}\Pi^{0}_{t}$ in TC2 model.
 The numerical results show that the cross section is very large
 (at the level of several tens fb), and over $10^3$ neutral toppion events
 can be produced in $e\gamma$ collision. With the large $Br(\Pi^0_t
 \rightarrow t\bar{c})$ and small cross section of $e^{-}\gamma
 \rightarrow e^{-}t\bar{c}$ in SM, $e^{-}\gamma\rightarrow e^{-}\Pi^0_t
 \rightarrow e^{-}t\bar{c}$ provide us the best channel to search for
 neutral toppion. On the other hand, the cross section of $e^{-}\gamma\rightarrow
e^{-}\Pi^0_t$ is about one order of magnitude larger than those of
the production processes of toppion-like particles in SM and MSSM.
Therefore, the process $e^{-}\gamma\rightarrow e^{-}\Pi^0_t$
provides us a unique way to distinguish TC2 model from other
models.




\newpage
\begin{figure}[h]
\begin{center}
\epsfig{file=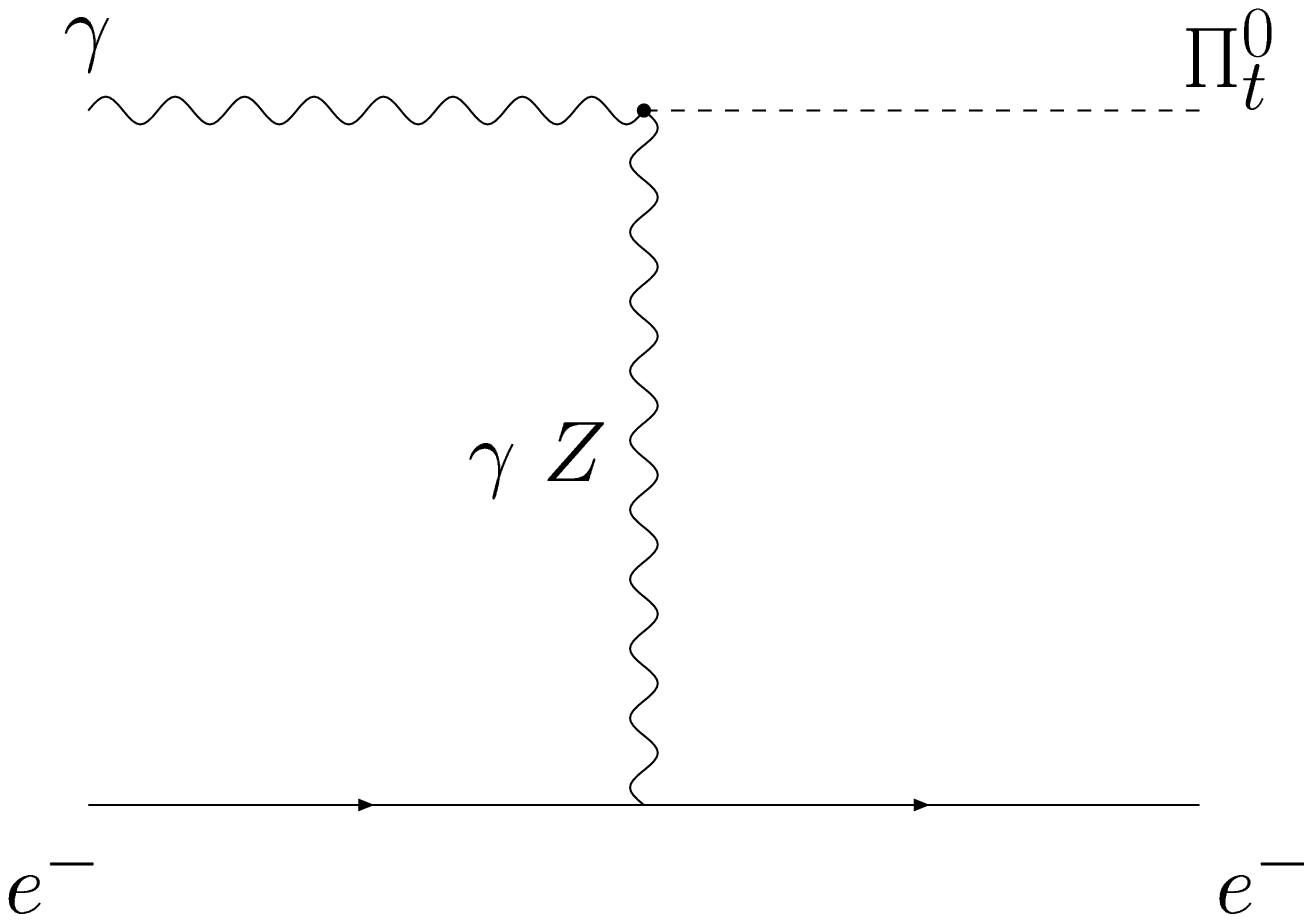,width=250pt,height=200pt} \vspace{-3.5 cm}
\caption{ The Feynman diagrams of the process
    $e^{-}\gamma\rightarrow e^{-}\Pi^{0}_{t}$.}
\label{fig1}
\end{center}
\end{figure}
\vspace{3 cm}

\begin{figure}[h]
\begin{center}

 \epsfig{file=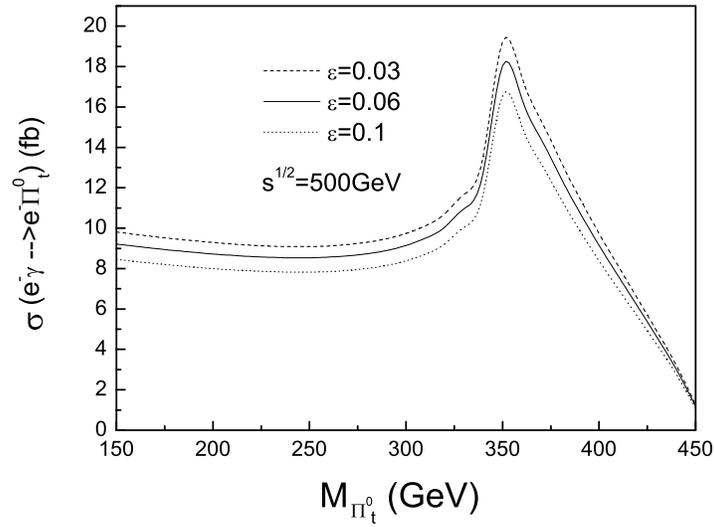,width=300pt,height=230pt}
\caption{The cross section of $e^{-}\gamma\rightarrow
e^-\Pi^{0}_{t}$ versus toppion mass $M_{\Pi}$(150-450 GeV) for
$\sqrt{s}=500$ GeV and $\varepsilon=0.03$(dash
line),$\varepsilon=0.06$ (solid
 line),$\varepsilon=0.1$(dot line), respectively }
\label{fig2}
\end{center}
\end{figure}

\begin{figure}[h]
\begin{center}
\epsfig{file=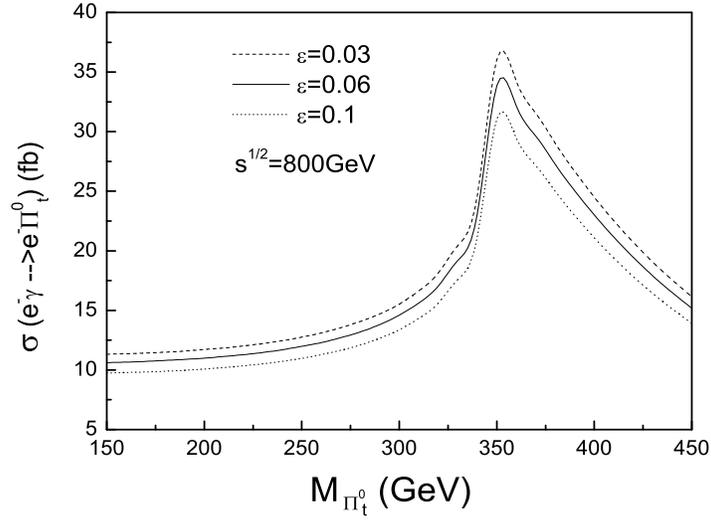,width=300pt,height=230pt} \caption{The same
plots as Fig.2 for $\sqrt{s}=800$.} \label{fig3}
\end{center}
\end{figure}

\begin{figure}[h]
\begin{center}
\epsfig{file=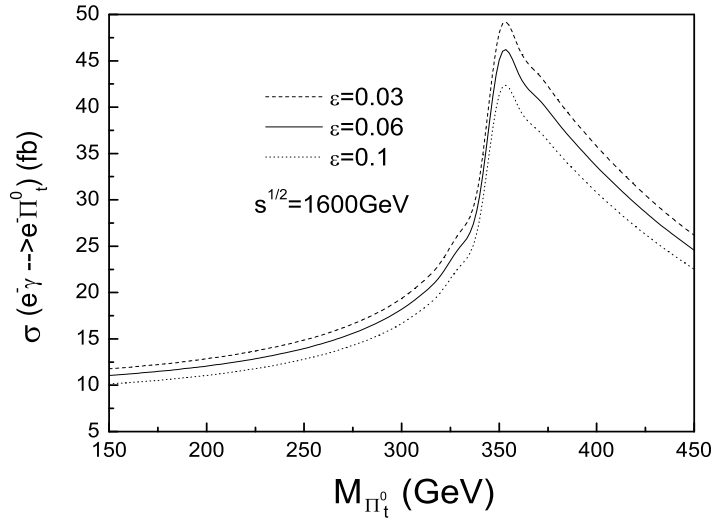,width=300pt,height=230pt} \caption{The
same plots as Fig.2 for $\sqrt{s}=1600$ GeV.} \label{fig5}
\end{center}
\end{figure}

\end{document}